\documentclass[article,balancelastpage,twocolumn,prl]{revtex4}%
\usepackage{makeidx}
\usepackage{amssymb}
\usepackage{dcolumn}
\usepackage{graphicx}
\usepackage{acronym}
\usepackage{amsmath}
\usepackage{amsfonts}%
\ifx\pdfoutput\relax\let\pdfoutput=\undefined\fi
\newcount\msipdfoutput
\ifx\pdfoutput\undefined\else
\ifcase\pdfoutput\else
\ifx\paperwidth\undefined\else
\ifdim\paperheight=0pt\relax\else\pdfpageheight\paperheight\fi
\ifdim\paperwidth=0pt\relax\else\pdfpagewidth\paperwidth\fi

\begin{document}
\title{The Dirac equation and anomalous magnetic moment}
\author{B. V. Gisin }
\affiliation{E-mail: borisg2011@bezeqint.net}
\date{\today }

\begin{abstract}
The idea of determining the magnetic resonance condition in a rotating frame of reference is being studied. In this frame the rotating magnetic field should be at rest. 
3D transformation that determines the transition to the frame is found.
The Dirac equation in a rotating electromagnetic field after such a transition is obtained. 
The magnetic resonance condition is determined by the stationarity of solutions. 
The connection between g-factor in the anomalous magnetic moment and a constant in 3D 
transformation is established. 

\end{abstract}

\maketitle

\section{Introduction}
Rotating magnetic fields in quantum mechanics are of
particular interest, especially for problem of the magnetic moment. This moment is measured in the state
of magnetic resonance. In measurements, the magnetic
field usually consists of a strong constant and a weak
oscillating field.

The rotating field is the most acceptable from the viewpoint of symmetry, but the creation of such a field is
technically more complicated than an oscillating field.
However the field can be represented as the sum of two
opposite rotating fields. The effect of one from them is
negligible and can be excluded from consideration.

It is well known that the Pauli equation is a non-relativistic approximation of the Dirac equation. The magnetic moment in the Pauli equation is equal to the Bohr
magneton. Measurements have shown that the magnetic
moment is different from this value.

The description of magnetic resonance is associated
with a transition in a rotating frame and the corresponding transformation of the wave function and coordinates.
However, for the Pauli equation, such a transition is accompanied only by a transformation of the wave function. This is due to the Newtonian non-relativistic nature
of time.

Note that a rotating field, especially a plane circularly
polarized electromagnetic wave, differs from mechanical
rotation. The rotating field has an axis of rotation at
every point perpendicular to the direction of propagation,
whereas mechanical rotation has only one such axis.

Below, a 3D transformation applicable to the first type
is found for differentials of the angle, distance along the
axis and time. The criterion for this case is the independence of angle in rotating frame from a distance perpendicular to the axis of rotation, because such an axis
exists at every point, and the dependence does not make
sense.

Some aspects of "point rotation" are described in \cite{pr}.

3D transformation is used for the Dirac equation in
a field of traveling circularly polarized electromagnetic
wave and constant magnetic field.

The paper shows that in rotating frame exact stationary solutions of Dirac's equation exist. The magnetic resonance condition is determined by the existence of such solutions. In
the non-relativistic approximation this condition exactly
coincides with the classical one.
\section{The Dirac equation}

Consider Dirac's equation
\begin{equation}
\lbrack-i\hbar\frac{\partial}{\partial t}+c\mathbf{\alpha}%
(\mathbf{\mathbf{p}}-\frac{e}{c}\mathbf{\mathbf{A}})+\beta mc^{2}]\Psi=0,
\label{D}%
\end{equation}

The potential $\mathbf{A}$ describes the traveling circularly polarized
electromagnetic wave propagating along the constant magnetic field $H_{3}$.%
\begin{align}
A_{1}  &  =\frac{1}{2}H_{3}y+\frac{1}{k}H\cos(\Omega t-kz),\label{Ax}\\
A_{2}  &  =-\frac{1}{2}H_{3}x+\frac{1}{k}H\sin(\Omega t-kz), \label{Ay}%
\end{align}
$k=2\pi/\lambda, \lambda, \Omega=kc$ are the propagation constant,
wavelength, angular frequency, respectively. For definiteness, 
the charge $e$ is assumed to be negative.

We denote the initial Dirac's operator in the laboratory frame as
\begin{equation}
D_0=-\hbar c(\frac{\partial}{c\partial t} +%
\alpha_{k}\frac{\partial}{\partial x_k})+\beta mc^{2},\;\;\;\label{D0}%
\end{equation}
and operator associated with the magnetic field as
\begin{equation}
\eta_0=\frac{1}{2}\alpha_2 H_{3}r\exp(\alpha_1\alpha_2\varphi)-
\alpha_1 H\exp[\alpha_1\alpha_2(\Omega t-kz)]. \label{A1}\\
\end{equation} 

Charged particles in such an electromagnetic field undergo a 
complex motion. Using the manipulations below, we find a rotating 
frame with a static magnetic field directed along the axis of rotation.

The field has cylindrical symmetry, so the cylindrical coordinates $r, \varphi, z, t$ are most suitable.
The change in Dirac's equation is achieved by passing to this coordinates and multiplying the equation by operator $P_c$
\begin{align}
P_c & =\exp(\frac{1}{2}\alpha_{1}\alpha_{2}\varphi),\: \Psi_c=P_c\Psi, \label{Pc}\\
D_c & =P_c D_0\tilde{P}_c, \:\: \eta_c=P_{c}\eta_0\tilde{P}_{c}. \label{etc}%
\end{align}

Here and below the tilde over operators $P_k$ corresponds to change sign in the exponent.
 
Operator $D_{c}(t,r,\varphi,z)$ becomes
\begin{equation}
-i\hbar c\left[\frac{\partial}{c\partial t} +%
\alpha_1\frac{\partial}{\partial r} +\alpha_1\frac{1}{2r}+%
\alpha_2\frac{\partial}{r\partial\varphi}+%
\alpha_3\frac{\partial}{\partial z}\right]+\beta mc^2, 
\end{equation}
and $\eta_c$ turns in 
\begin{align}
\eta_c = \alpha_2\frac{e}{2}H_{3}r+ \frac{e}{k}H(-\alpha_1\cos\bar{\varphi}+\alpha_2\sin\bar{\varphi}). \label{Eec}
\end{align}
The variable $\bar{\varphi}$ in Eq. (\ref{Eec}) is
\begin{equation}
\bar{\varphi} =\varphi+kz-\Omega t.  \label{Phb}%
\end{equation} 

\section{3D Transformation for rotating frames}

Line elements of cylindrical coordinates are orthogonal
similarly to Cartesian coordinates.

Dirac’s equation is invariant under rotations of elements of cylindrical coordinates $rd\varphi, dz, cdt$ by constant
angles.

It is well known that a 3D transformation can be
obtained by sequentially rotating three pairs of the elements. This invariance for Dirac's operator without the derivative in respect to $r$ can be extended to rotation by
angles depending on $r$. However, new terms should arise in complete
Dirac’s equation.

Emphasize, line elements $rd\varphi, cdt$ and also other pairs converted into elements of non-inertial frames. Appearance of new terms is, in a sense, an analog of centrifugal forces in mechanics.

Sequentially rotations of all coordinate pairs involve
the change of parameter $\eta$ together with new terms,
whereas the form of Dirac’s operator $D_c$ is kept.

From classical physics it is well known that in a strong
magnetic field the electron rotates around the direction
of field. It can be compensated by the rotation in plane
of line elements $(rd\varphi, cdt)$, which is described by the
Lorentz transformation
\begin{align}
& rd\varphi^{\prime} =rd\varphi\cosh\Phi-cdt\sinh\Phi,\label{Tp1}\\
& \text{ }c dt^{\prime} =-rd\varphi\sinh\Phi+c dt\cosh\Phi,\label{Tt1}\\
& \tanh\Phi =r\Omega/c, \label{tanh}%
\end{align}
where $r\Omega$ is the line velocity on the circle of radius $r$. The
sign in equality (\ref{tanh}) must be consistent with the rotating
magnetic field.

The denominator 
of $\sinh\Phi$ and $\cosh\Phi$ is $\sqrt{1-\Omega^{2}r^{2}/c^{2}}$.
It means that the maximal value of $r$ is bounded:
\begin{equation} 
 r\leq\ \frac{\lambda}{2\pi}. \label{rM}
\end{equation}
For small velocity $|\Omega r|\ll c$ the transformations turns into a
non-relativistic rotation: $d\varphi^{\prime}=d\varphi-\Omega dt,$ $dt^{\prime}=dt$

The transition to primed coordinates in Dirac's operator is
 realized with help of derivatives transformation
\begin{align}
\frac{\partial}{r\partial\varphi}  &  =\cosh\Phi\frac{\partial}{r\partial
\varphi^{\prime}}-\sinh\Phi\frac{\partial}{c\partial t^{\prime}},\label{dp1}\\
\text{ \ }\frac{\partial}{c\partial t}  &  =-\sinh\Phi\frac{\partial
}{r\partial\varphi^{\prime}}+\cosh\Phi\frac{\partial}{c\partial t^{\prime}}.\label{dt1}%
\end{align}
The derivatives may easily be found from (\ref{Tp1}) and (\ref{Tt1}). 
  
Also Dirac's equation should be multiplied by the operator $P_1$.
\begin{align}
& P_1 =\exp(\frac{1}{2}\alpha_2\Phi),\: \Psi_1=\tilde{P}_1\Psi, \label{P1}\\
& D_1 =P_1 D_c P_1=D_c(t',r,\varphi',z)-i\frac{1}{2}\hbar c\alpha_1\alpha_2\Phi_{,r},  \label{De1} \\
& \eta_1 =P_1\eta_c P_1. \label{eta1} %
\end{align}
where the comma means differentiation $\Phi_{,r}=\partial{\Phi}/\partial{r}$. Terms with derivatives in respect to $r$, similarly (\ref{De1}), should arise after each new rotation. 

The above manipulation compensate the rotation at
frequency $\Omega$. However, there are magnetic and electric
fields of circularly polarized electromagnetic wave which
is not so strong as $H_3$. The fields contribute to the rotation, as follows from classical physics. 

The magnetic field
adds motion perpendicular to the plane of rotation, the
electric field adds motion along the direction of rotation.
For compensation of that we must add a rotation in the
plane of line elements $(dz; rd\varphi')$:
\begin{align}
dz' &  =dz\cos\Phi_{1}-rd\varphi^{\prime}\sin\Phi_{1},\label{Tz2}\\
\text{ \ }rd\tilde{\varphi}  &  =dz\sin\Phi_{1}+rd\varphi^{\prime}\cos\Phi
_{1}, \label{Tp2}%
\end{align}
This rotation should be accompanied by the multiplication of Dirac's equation by the operator $P_2$.
\begin{align}
& P_2=\exp(\frac{1}{2}\alpha_2\alpha_3\Phi_1),\quad \Psi_2=P_2\Psi_1, \\
& D_2=P_2 D_1\tilde{P}_2, \quad \eta_2=P_2\eta_1\tilde{P_2}. \label{eta2} 
\end{align}

We also must add the Lorentz transformation of line elements $ dz^{\prime},dt^{\prime}$
\begin{align}
d\tilde{z}  &  = dz^{\prime}\cosh\Phi_{2}-c dt^{\prime}\sinh\Phi_{2}%
,\label{Tz3}\\
c d\tilde{t}  &  =-dz^{\prime}\sinh\Phi_{2}+c dt^{\prime}\cos\Phi_{2},
\label{Tt3}%
\end{align}
and corresponding change of $D_2, \Psi_2, \eta_2$ with help of operator $P_3$.
\begin{align}
& P_3=\exp(\frac{1}{2}\alpha_3\Phi_2),\quad \Psi_3=P_3\Psi_2, \\
& D_3=P_3 D_2\tilde{P}_3, \quad \eta_3=P_3\eta_2\tilde{P_3}. \label{eta23} 
\end{align}
This transformation allows us to correct momentum along the new $\tilde{z}$ axis

3D transformation, as the coordinate dependence
$(rd\tilde{\varphi}, d\tilde{z}, cd\tilde{t})$ on $(rd\varphi, dz, cdt)$, can be obtained after excluding the primed coordinates from above rotations.

The angle $\Phi$ in 3D transformation is specified by (\ref{tanh}). $\Phi_1$ is defined by the condition of speed of light constancy 
\begin{align}
\frac{d\tilde{z}}{d\tilde{t}}=c, \quad \mathrm{if} \quad \frac{dz}{dt}=c.
\end{align}
From this condition we get the connection between $\Phi_{1}$ and $\Phi$%
\begin{equation}
\sin\Phi_{1}=\tanh\Phi=\Omega r/c, \label{sc}%
\end{equation}

The angle $\Phi_{2}$ is determined from following requirement.
The 3D transformation and $\tilde{\eta}$ have singularity of the shape
$1/\sqrt{1-r^{2}\Omega^{2}/c^{2}}$ and $1/(1-r^{2}\Omega^{2}/c^{2})$ at
$|\Omega r|\rightarrow c.$ The singularity vanishes, if 
$\Phi_{2}$ is defined from the condition%
\begin{equation}
\Phi_{2,r}=\Phi_{,r}\sin\Phi_{1},\text{ }\exp\Phi_{2}=\frac{\varkappa}%
{\sqrt{1-\Omega^{2}r^{2}/c^{2}}}, \label{f2}%
\end{equation}
moreover $D_3$ and $\eta_3$ are simplified.

After simplification
the 3D transformation may be written as follows%
\begin{align}
d\tilde{\varphi}  &  =d\varphi+kdz-\Omega dt,\label{tphi}\\
d\tilde{z}  &  =-\frac{kr^{2}}{\varkappa}d\varphi+\varkappa_{22}%
dz+\varkappa_{23}cdt,\label{tz}\\
cd\tilde{t}  &  =-\frac{kr^{2}}{\varkappa}d\varphi+\varkappa_{32}%
dz+\varkappa_{33}cdt. \label{tt}%
\end{align}
The reverse transformation is
\begin{align}
d\varphi  &  = d\tilde{\varphi}-\frac{k}{\varkappa}d\tilde{z}+\frac{\Omega}{\varkappa} d\tilde{t},\label{rphi}\\
dz  &  = kr^{2}d\tilde{\varphi}+\varkappa_{22}%
d\tilde{z}-\varkappa_{32}cd\tilde{t},\label{rtz}\\
cdt  &  = kr^{2}d\tilde{\varphi}-\varkappa_{23}%
d\tilde{z}+\varkappa_{33}cd\tilde{t}. \label{rt}
\end{align}
$\varkappa$ is the dimensionless constant. 

It is shown below that $\varkappa$ 
is nothing but the $g$ factor in the expression of anomalous magnetic moment. 

$\varkappa_{kl}$ is
defined as follows
\begin{align}
\varkappa_{22}  &  =\frac{1}{2\varkappa}(1+\varkappa^{2}-\frac{1}{c^{2}}%
\Omega^{2}r^{2}),\label{C22}\\
\varkappa_{23}  &  =\frac{1}{2\varkappa}(1-\varkappa^{2}+\frac{1}{c^{2}}%
\Omega^{2}r^{2}),\label{C23}\\
\varkappa_{32}  &  =\frac{1}{2\varkappa}(1-\varkappa^{2}-\frac{1}{c^{2}}%
\Omega^{2}r^{2}),\text{ }\label{C32}\\
\varkappa_{33}  &  =\frac{1}{2\varkappa}(1+\varkappa^{2}+\frac{1}{c^{2}}%
\Omega^{2}r^{2}). \label{C33}%
\end{align}

The dependence $\tilde{\varphi}(\varphi,z,t)$ in (\ref{tphi}) can be 
written in integral form, that is 
$\bar{\varphi}=\tilde{\varphi}$, unlike variables $d\tilde{z}, d\tilde{t}$. 
It means that the criterion noticed in Introduction is fulfilled.

It is easy to see that the determinant of 3D transformation as well as 
its reversal is 1. This is important, since the determinant defines 
Jacobian and volume element in different coordinate systems. 

Also note that the 3D interval $r^{2}d\varphi^{2}+dz^{2}-c^{2}dt^{2}$ and 3D differential form $\partial^2/r^2\partial\varphi^2+\partial^2/\partial{z^2}-\partial^2/c^2\partial{t^2}$ are kept under the 3D transformation. Along with this, the 4D interval 
\begin{align}
 ds^2=dr^2+r^{2}d\varphi^{2}+dz^{2}-c^{2}dt^{2} \label{ds} %
\end{align}
is also preserved.
 
This is an important and far-reaching conclusion.  

Noteworthy, all the three above rotation are transformations at "constant $r$". This is possible since differentials of line elements are used and the relationship between them is built on the basis of general physical laws. In transformations, the change of $r$ is the second order of smallness and can be neglected. 

However, in the final reversal of Dirac's equation from the cylindrical to Cartesian coordinates this dependence must be taken into account as it is in the initial transformation from the Cartesian to cylindrical coordinates. 

We use in the reversal variables $\tilde{x}, \tilde{y}$ and operator $P_{r}$
\begin{align}
& \tilde{x}=r\cos\tilde{\varphi}, \quad \tilde{y}=r\sin\tilde{\varphi} \\
& P_{r}=\exp(-\frac{1}{2}\alpha_{1}\alpha_{2}\tilde{\varphi}), \quad  \Psi_c=P_r\Psi_2, \\
& D_c(\tilde{t},r,\tilde\varphi,\tilde{z})=P_{r}D_3\tilde{P_{r}},\quad \eta_c =P_{r}\eta_3\tilde{P_{r}}. %
\end{align} 

As result Dirac's equation takes the form 
\begin{align}
& \left(D_0 (\tilde{t},\tilde{x},\tilde{y},\tilde{z})- 
\alpha_{1}\frac{e}{k}H-\alpha_{1}\frac{e}{2}H_{3}\tilde{y} +\alpha_{2}\frac{e}{2}H_{3}\tilde{x}\right)\tilde{\Psi}+ \nonumber \\
& (1-\alpha_{3})\left({-i\alpha_{1}\alpha_{2}\frac{\hbar\Omega}
{2\varkappa}+\frac{e}{\varkappa}H\tilde{y}+\frac{\Omega e}{2c\varkappa}
H_{3}r^{2}}\right)\tilde{\Psi}=0, \nonumber
\end{align}
where $\tilde{\Psi} =P_{r}\tilde{P_{3}}P_{2}\tilde{P_{1}}P_c\Psi$.

Frequency of the magnetic field $\Omega$ is contained in all expressions corresponding to the rotating frame only in the form $\Omega/\varkappa$. This allows us to assume that the shape will be kept in the condition of magnetic resonance.

\section{Solutions in rotating frame}

Before this section, the concrete form of Dirac's matrices was not used. 
It is well known that the matrices are connected only by the condition 
\begin{equation}
\alpha_{\mu}\alpha_{\nu}+\alpha_{\nu}\alpha_{\mu}=\delta_{\mu\nu}.\label{mn} 
\end{equation}

For the given problem it is more convenient to exchange the conventional 
matrices 
\begin{equation}
\alpha_{3} \leftrightarrow \beta. \label{c}%
\end{equation}

After that multiply Dirac's equation by
$\frac{1}{2}(1+\beta)$ and $\frac{1}{2}(1-\beta)$ and take
into account that four-component spinor $\tilde{\Psi}$ can
be represented as two two-component spinors. Denote the 
"upper" and "lower" part of $\tilde{\Psi}$ as $\Psi_1$ and $\Psi_2$.

Dirac's equation then breaks down into two equations
\begin{align}
& Q\Psi_2-i\hbar\left(\frac{\partial}{\partial\tilde{t}}+ 
c\frac{\partial}{\partial\tilde{z}}\right)\Psi_1=0, \label{Dp1} \\
& Q\Psi_1-i\hbar\left(\frac{\partial}{\partial\tilde{t}}- 
c\frac{\partial}{\partial\tilde{z}}\right)\Psi_2+Q_{s}\Psi_2=0, \label{Dm1} 
\end{align}
where
\begin{align}
Q  & =(\sigma_1 Q_{1}+\sigma_2 Q_2+\sigma_3 mc^2) \label{Q} \\
Q_{s} & =\sigma_3\frac{\hbar\Omega}{\varkappa} +2\frac{e}{\varkappa}H\tilde{y}+\frac
{e\Omega}{c\varkappa}H_{3}r^{2},\label{Qs}\\
Q_{1} &  =-ic\hbar\frac{\partial}{\partial\tilde{x}}-\frac{e}{k}H
-\frac{e}{2}H_{3}\tilde{y},\label{Q1}\\
Q_{2} &  =-ic\hbar\frac{\partial}{\partial\tilde{y}}+
\frac{e}{2}H_{3}\tilde{x}, \label{Q2}%
\end{align}
and $\sigma_{k}$ is the Pauli matrices.

Multiply Eq. (\ref{Dm1}) by
\[ i\hbar\left(\frac{\partial}{\partial\tilde{t}}+ 
c\frac{\partial}{\partial\tilde{z}}\right) \]
and exclude $\Psi_1$ with help of Eq. (\ref{Dp1}), then we obtain equation of the second order for $\Psi_2=0$
\begin{align}
& Q^{2}\Psi_2+ \hbar^2(\frac{\partial^2}{\partial\tilde{t^2}}- 
c^2\frac{\partial^2}{\partial\tilde{z^2}})\Psi_2+ \nonumber \\
& Q_{s}i\hbar(\frac{\partial}{\partial\tilde{t}}+ 
c\frac{\partial}{\partial\tilde{z}})\Psi_2=0, \label{Dm2} 
\end{align}

For stationary states,
\begin{equation}
i\hbar\frac{\partial}{\partial\tilde{t}}\tilde{\Psi}=\tilde{E}{\Psi}, \quad 
ic\hbar\frac{\partial}{\partial\tilde{z}}\tilde{\Psi}=-c\tilde{p}{\Psi},
\label{Ep}
\end{equation} 
where $\tilde{E}$ and $\tilde{p}$ is "energy" and "momentum" in the rotating frame, Eq. (\ref{Dm2}) becomes 
\begin{align} 
& [\sigma_3\hbar ceH_3 +\sigma_3\frac{\hbar\Omega}{\varkappa}(\tilde{E}-c\tilde{p})+ \frac{e^2}{k}HH_3\tilde{y}+
2ic\hbar\frac{e}{k}H\frac{\partial}{\partial\tilde{x}}- \nonumber\\                                         
& \hbar^{2}c^2(\frac{\partial^2}{\partial\tilde{x^2}}+ \frac{\partial^2}{\partial\tilde{y^2}})-ic\hbar eH_3(\tilde{x}\frac{\partial}{\partial\tilde{y}}-
\tilde{y}\frac{\partial}{\partial\tilde{x}})+(\frac{e}{k}H)^2- \nonumber \\
& \tilde{E^2}+c^2\tilde{p^2}+m^2c^4+(\frac{e}{2}rH_3)^2+
\frac{e\Omega}{c\varkappa}H_{3}r^{2}(\tilde{E}-c\tilde{p})+ \nonumber \\ 
& 2\frac{e}{\varkappa}H\tilde{y}(\tilde{E}-c\tilde{p})]\Psi_2 = 0. \label{Eqfs} %
\end{align}

We are looking solutions of this equation in form
\begin{align}
\Psi_2=\exp(D)\psi, \quad D=ar^2+q_1\tilde{x}+q_2\tilde{y}.
\label{Dp}
\end{align}

The parameter $a$ is inserted for compensation term with $r^2$ in Eq. (\ref{Eqfs}). This parameter is determined by equating the factor at $r^2$ to zero. 
\begin{equation}
a=\pm i\frac{\sqrt{3}H_3}{2c\hbar}. \label{a}
\end{equation} ,

Two components of the wave function $\Psi_2$ are described by independent equations. Equations differ by constant term with matrix $\sigma_3$. If the term is equal zero, then energy $\tilde{E}$ is  the same for these components and states are stationary, in opposite case they are non-stationary. Below we consider first case. 

Equate the sum of terms with $\sigma_3$ to zero
\begin{align}
ceH_3=-\frac{\Omega}{\varkappa}(\tilde{E}-c\tilde{p}).
\label{mr}
\end{align}   
This is the condition of magnetic resonance. 

Eq. (\ref{Eqfs}) may be rearranged so that variables can be separated.

Use new variables
\begin{align}
 \tilde{x}'=\tilde{x}+b_1, \quad \tilde{y}'=\tilde{y}+b_2.  %
\end{align}

Substitute the variables in Eq. (\ref{Eqfs}). We can find $b_1,\: b_2$, equating to zero the sum of factors at $x',\: y'$ respectively; equating the sum of factors at first derivatives in respect to $x'$ and $y',$ we find $q_1$ and $q_2$.
\begin{align}
b_1=0, \;  b_2=\frac{H}{kH_3}, \quad
& q_1=i\frac{e}{2kc\hbar}H, \; q_2=a\frac{H}{kH_3}. %
\end{align} 

In the new cylindrical coordinates
\begin{align}  
\tilde{x}'=\rho\cos\phi, \quad \tilde{y}'=\rho\sin\phi,   
\end{align}
the equation allows separation variables 
\begin{align} 
&  [\hbar^{2}c^2(\frac{\partial^2}{\partial\rho^2}+ \frac{1}{\rho}\frac{\partial}{\partial\rho}+ \frac{\partial^2}{\rho^2\partial\phi^2}+ 2a\rho\frac{\partial}{\partial\rho})+ic\hbar eH_3\frac{\partial}{\partial\phi} \nonumber \\
& +2a\hbar^{2}c^2+2\frac{e^2}{k^2}H^2+ \tilde{E^2}-c^2\tilde{p^2}-m^2c^4]\psi = 0. \label{Ef5} %
\end{align}

Obviously, the operator $-i\partial/\partial\phi$ commutes with all operators of the equation. Accordingly to rules of quantum mechanics we can determine eigenvalues of this operator and dependence of the wave function on $\phi$ 
\begin{align}
& -i\frac{\partial}{\partial\phi}\psi=\mathfrak{m}\psi, \quad \psi=\exp(i\mathfrak{m}\phi)\psi_{\rho}(\rho), %
\end{align}
where $\mathfrak{m}$ is an integer.

Let $\psi=\rho^\mathfrak{m}f(\rho)$. Eq. (\ref{Ef5}) turns into 
\begin{align} 
&  [\hbar^{2}c^2(\frac{d^2}{d\rho^2}+ \frac{2\mathfrak{m}+1}{\rho}\frac{d}{d\rho}+ 2a\rho\frac{d}{d\rho}) 
-c\hbar eH_3\mathfrak{m}+2\frac{e^2}{k^2}H^2 \nonumber \\
& +2a\hbar^{2}c^2(\mathfrak{m}+1)+ \tilde{E^2}-c^2\tilde{p^2}-m^2c^4]f = 0. \label{Ef6} %
\end{align}

A problem of this equation is the imaginary parameter $a$. The availability of $a$ can lead to complex values of $\tilde{E}$. 

A simplest example of the solution with real $\tilde{E}$ is at $\mathfrak{m}=-1$ and a constant $f$, \[\psi=\frac{1}{\rho}\exp(-i\phi).\] The real value of $\tilde{E}$ is determined from characteristic equation 
\begin{equation}
  \tilde{E^2}=c^2\tilde{p^2}+m^2c^4-c\hbar eH_3 -2\frac{e^2}{k^2}H^2, %
\end{equation}

In non-relativistic case $\tilde{p}\ll mc$, $\tilde{E}\approx mc^2$ the condition (\ref{mr}) exactly coincides with condition of the magnetic resonance with $\varkappa$ as $g$-factor 
\begin{equation}
\Omega=-g\frac{e}{mc}H_3.  
\label{mm}%
\end{equation}

Obviously, all parameters in this equation $(\omega, H_3, m, c)$ are measured in the laboratory frame. The sign in (\ref{mm}) depends on direction of the magnetic field $H_3$ and sense of rotation frequency $\Omega$. As noted in Section 2, $e=-|e|$.

The second two-component spinor $\Psi_1$ is easily calculated from Eq. (\ref{Dp1}).

\section{Conclusion}

The description of magnetic resonance is associated with the transition to a
rotating frame with a static magnetic field. In this frame the results of
magnetic resonance measurements should be interpreted.
However, such a transition in the framework of Pauli equation is incorrect
because of the Newtonian time in this equation. 

The transformation of
Dirac equation into a rotating frame can be carried out with help of 
3D transformation. The magnetic moment is determined from the requirement of solutions stationarity, because this moment rotates together with the frame. 

The condition of magnetic resonance, found in the paper, is described by an exact formula that can be used for experimental verification.

In this approach the magnetic moment is always anomalous. In non-relativistic approximation $g$-factor is equal to the constant $\varkappa$ in 3D transformation.
This approach does not lead to a specific value for the $g$-factor. The value
is determined by the properties of measured object and is calculated by
quantum field theory. 

The approach does not contradict the theory and it would be desirable to revise some foundations of this theory with respect to rotating fields.

In light of the above results and, in particular, the surprising conclusion that the 4D Euclidean metric is preserved under the 3D transformation, there arise sacramental questions: Is our space an object of point rotation? And what is frequency of its rotation? These questions is another reason for studying this problem, applicable to various field of physics.

\end{document}